\documentclass[english,aps,pra,twocolumn,showpacs,superscriptaddress]{revtex4}

\usepackage[T1]{fontenc}
\usepackage[latin9]{inputenc}
\usepackage{verbatim}
\expandafter\let\csname equation*\endcsname\relax
\expandafter\let\csname endequation*\endcsname\relax 
\usepackage{amsmath}
\usepackage{amssymb}
\usepackage{graphicx}
\usepackage{esint}
\usepackage{babel}
\usepackage{epstopdf}
\usepackage{upgreek}

\begin{document}

\title{Individual Addressing in Quantum Computation through Spatial
Refocusing}
\author{C. Shen}
\email{chaoshen@umich.edu}
\affiliation{Department of physics, University of Michigan, Ann Arbor, Michigan 48109, USA}
\affiliation{Center for Quantum Information, IIIS, Tsinghua University, Beijing, China}

\author{Z.-X. Gong}
\affiliation{Department of physics, University of Michigan, Ann Arbor, Michigan 48109, USA}
\affiliation{Joint Quantum Institute, University of Maryland, College Park, Maryland 20742, USA}

\author{L.-M. Duan}
\affiliation{Department of physics, University of Michigan, Ann Arbor, Michigan 48109, USA}
\affiliation{Center for Quantum Information, IIIS, Tsinghua University, Beijing, China}

\begin{abstract}
  Separate addressing of individual qubits is a challenging
  requirement for scalable quantum computation, and crosstalk between
  operations on neighboring qubits remains a significant source of
  error for current experimental implementations of multi-qubit
  platforms. We propose a scheme based on spatial refocusing from
  interference of several coherent laser beams to significantly reduce
  the crosstalk error for any type of quantum gates. A general
  framework is developed for the spatial refocusing technique, in
  particular with practical Gaussian beams, and we show that the
  crosstalk-induced infidelity of quantum gates can be reduced by
  several orders of magnitude with a moderate cost of a few correction
  laser beams under typical experimental conditions.
\end{abstract}

\pacs{03.67.Lx, 03.67.Ac, 32.80.Qk}
\maketitle

\section{Introduction}
Performing useful quantum computation and simulation in the presence
of unavoidable noise has been a goal long sought after. Many solid
steps have been taken on different physical platforms in the past
decade, demonstrating for small systems elementary quantum logic
\cite{logic}, simple algorithms
\cite{algo}, error
correction \cite{QEC} and quantum simulation
\cite{qsim}. While the celebrated error threshold
theorem \cite{nielsen_chuang} guarantees the fault tolerance of a
large scale quantum computer when each single operation error is
reduced below a certain limit, this threshold is very hard to satisfy
in a typical multi-qubit setting. To fully control the state evolution
of the quantum information processor, one needs to pinpoint any
individual qubit at will and manipulate it while keeping the others
intact. This is a stringent requirement for almost all physical
platforms. A lot of efforts have been devoted to the development of
individual addressing optical beam delivery and imaging systems
\cite{JSKim_mems, atom_imaging}. Assuming a Gaussian profile
of the beam, single qubit addressing typically requires the beam waist
to be much smaller than the inter-qubit spacing, which is half the
wavelength of the trapping laser in optical lattices and around one
micron in a linear trapped ion chain. So subwavelength focusing beyond
the diffraction limit is usually required and this makes it
experimentally very challenging.

There have already been many proposals and/or demonstrations in the
context of cold atoms in optical lattices \cite{saffman_opt_lett,
  dark_state, demon_dark_state, single_spin_lattice} and linearly
trapped ions \cite{KBrown_single_bit, KBrown_multi_bit}. To name a
few, interference of several Bessel beams were proposed to form a
pattern such that all but one atom locate at the nodes of laser
profile in \cite{saffman_opt_lett}; the sharp nonlinear atomic
response and position dependent dark states in an electromagnetically
induced transparency (EIT) setting was exploited to enable
subwavelength selectivity in various proposals \cite{dark_state} and
experimentally demonstrated very recently \cite{demon_dark_state};
single spin manipulation in an optical lattice with the combination of
a well focused level shifting beam and a microwave pulse was
demonstrated in \cite{single_spin_lattice}. The adaptation of
composite pulse refocusing technique widely used in nuclear magnetic
resonance \cite{NMR} and quantum information \cite{DD} to trapped ions
was considered for single-qubit operation \cite{KBrown_single_bit} and
two-qubit operation with a special form of interaction
\cite{KBrown_multi_bit}. Note the two-qubit correction scheme depends
on the physical operation being carried out and requires specific form
of controllable interaction, and does not reduce error for certain
gate realizations. 

Our approach is along the line of \cite{saffman_opt_lett} but in a
different setting. We propose and provide detailed analysis for a
scheme to reduce crosstalk error and achieve individual addressing
with several imperfectly focused laser beams. By applying an array of
beams centered at different qubits and controlling their relative
amplitudes, we can achieve quantum gates with ideal fidelity even when
the beam waist is comparable with or slightly larger than the
inter-qubit distance. A reduction of the crosstalk error by several
orders of magnitude can be achieved with only moderate increase of the
required laser power. The basic idea is reminiscent of the refocusing
in NMR, but works in the spatial domain using multiple beams instead
of in the time domain. So we call this technique spatial
refocusing. Unlike \cite{KBrown_multi_bit}, this technique is
universal and works for any quantum gate. We believe it is a valuable
addition to the existing toolbox of subwavelength addressing. 

\section{Mathematical formulation}
We consider an array of qubits with even spacing $a$ 
located at the positions $x_{i}$ $\left( i=1,2,\cdots ,N\right) $. The
laser beam used to manipulate the qubits is assumed to have a spatial
profile denoted by $g(x-x_{i})$ when it is centered at $x_{i}$. To
have individual addressing, normally we assume the laser is strongly
focused so that $g(x_{j}-x_{i})\rightarrow 0$ for any $j\neq i$ (i.e.,
$%
g(x_{j}-x_{i})=\delta _{ij}$). It remains experimentally challenging
to achieve this condition in multi-qubit quantum computing platforms
where the spacing $a$ needs to be small to have sufficiently strong 
interaction. Here, instead of strong focusing, we assume that the
laser beams applied to different qubits have relative coherence. To
address a single qubit, say qubit $i$ at position $x_{i}$, instead of
just shining this qubit with $%
g(x-x_{i})$, we apply a number of identical beams centered on its
nearby qubits with relative amplitudes denoted by
$f(x_{j}-x_{i})$. The total effective laser profile is then the
convolution%
\begin{equation}
G(x-x_{i})=\sum_{j}g(x-x_{j})f(x_{j}-x_{i}).
\end{equation}%
For a given $g(x-x_{i})$, we want to find an envelop function
$f(x_{j}-x_{i}) $ to make $G(x_{j}-x_{i})\rightarrow 0$ for any $j\neq
i$. It is desirable that $f(x_{j}-x_{i})$ is fast decaying so that in
practice we can cut off $j$ in the summation of Eq. (1) and apply
laser beams to only a few of its neighbors. If we take the
normalization $g(0)=G(0)=1$, $%
f(0)$ then determines the relative increase of the required laser
light amplitude, which is desired to be moderate for practical applications.

The solution depends on the laser profile $g(x-x_{i})$. To show that
the idea works, first we look at a toy model by assuming $g(x-x_{i})$
given by an exponential decay $g(x-x_{i})=e^{-\alpha \left\vert
    x-x_{i}\right\vert }$. In this case, two correction beams applied
to its nearest neighbors $x_{i-1}$ and $x_{i+1}$ perfectly cancel the
residue laser amplitude for all the qubits $j\neq i$. To see this,
let us take $%
f(0)=\beta _{0}$, $f(x_{j}-x_{i})=\beta _{1}$ for $j=i\pm 1$, and all
other $%
f(x_{j}-x_{i})=0$. If we choose $\beta _{0}=\left( 1+\lambda
  ^{2}\right) /\left( 1-\lambda ^{2}\right) $ and $\beta _{1}=-\lambda
/\left( 1-\lambda ^{2}\right) $, where $\lambda \equiv e^{-\alpha a}$,
we immediately have $%
G(x_{j}-x_{i})=\delta _{ij}$. The required increase of the laser power
$%
f(0)=\left( 1+\lambda ^{2}\right) /\left( 1-\lambda ^{2}\right) $ is
moderate even when the original laser profile $g(x-x_{i})$ has a
significant residue amplitude $\lambda =e^{-\alpha a}$ on the
neighboring qubits.

For a general laser profile $g(x-x_{i})$, if the number of qubits is large
or if the envelop function $f(x_{i})$ is fast decaying so that the boundary
condition is irrelevant, we can formally solve Eq. (1) by assuming the
periodic boundary condition for the array. In this case, we can take a
discrete Fourier transformation of Eq. (1), which yields $g(k)f(k)=G(k)$. As
the target profile $G(x-x_{i})$ needs to be a $\delta $-function, $G(k)=1$,
and a formal solution of Eq. (1) is
\begin{equation}
f(x_{j}-x_{i})=\frac{1}{N}\sum_{k}\frac{1}{g(k)}e^{ik(x_{j}-x_{i})/a},
\end{equation}%
where the summation is over $k=\pi n/N$ with $n=-N/2, -N/2+1,\cdots ,N/2$. In the
limit of large $N$, $f(x_{j}-x_{i})\approx \left( 1/2\pi \right) \int_{-\pi
}^{\pi }dk\left[ 1/g(k)\right] e^{ik(x_{j}-x_{i})/a}$.

\begin{figure}[tbp]
\centering     
\includegraphics[width=0.45\textwidth]{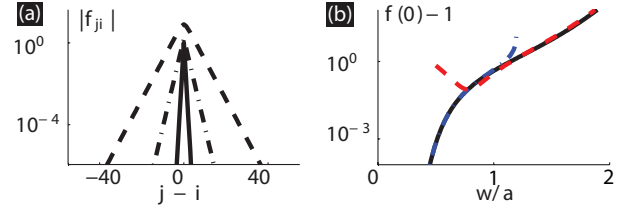}
\caption{(Color online) (a): Envelope function $f_{ji}$ under different Gaussian beam waist ($w/a = 1.5, 1.0, 0.5$ for curves from top to bottom), calculated for a
homogeneous chain of $401$ qubits. Near the center $j-i=0$, $f_{ji}$ has
co-existing components with different decay constants so $%
\left|f_{ji}\right|=(-1)^{j-i}f_{ji}$ deviates from a straight line on the
log plot. Only a few lattice sites away, $\left|f_{ji}\right|$ straightens
and the slope matches that of $\protect\gamma^{|j-i|}$ precisely. 
(b): Amplitude $f(0)$ versus the beam waist $w/a$. For visibility $f(0)-1$ is plotted. Black solid
line is from numerical exact integration of Eq.(2) and the blue dash-dot
(red dashed) line is from the analytic approximation $f(0)=1/\protect\sqrt{%
1-4\protect\gamma^{2}}$ ($f(0)=\frac{2}{\protect\pi^{5/2}w^{3}}e^{\protect\pi%
^{2}w^{2}/4}$), valid for the region $w/a\lesssim 1$ ($w/a\gtrsim 1$). }
\label{fig:f0_fx_two_panel}
\end{figure}

Now we apply this formalism to practical Gaussian beams, for 
which $%
g(x-x_{i})=\exp\left[ -\left( x-x_{i}\right) ^{2}/w^{2}\right] $,
where $w$ characterizes the width of the beam. The discrete Fourier
transformation of $%
g(x-x_{i})$ gives

\begin{equation}
g(k)=\sum_{n\in Z}\exp [-\left( na\right) ^{2}/w^{2}]\exp (-ikn)=\theta
_{3}(k/2,\gamma )
\end{equation}%
where $\gamma \equiv e^{-a^{2}/w^{2}}<1,$ and $\theta _{3}(z,q)\equiv
1+2\sum_{n=1}^{\infty }q^{n^{2}}\cos(2nz)$ is the Jacobi elliptic
function.  We can do a series expansion with $\gamma $, and up to the
order of $\gamma ^{2}$, $g(k)\approx 1+2\gamma
\cos(k)+\mathcal{O}(\gamma ^{4})$ and $%
f(x_{j}-x_{i})\approx \left( 1+2\gamma ^{2}\right) \delta _{ij}-\gamma
\delta _{i\pm 1,j}+\gamma ^{2}\delta _{i\pm 2,j}$. One can see that
the envelop function $f(x_{j}-x_{i})$ decays exponentially by the
factor $%
-\gamma $ as one moves away from the target qubit. This result holds
in general. To show this, we write Eq. (1) into a matrix form $%
\sum_{j}M_{nj}f_{ji}=\delta _{n,i}$, denoting $x$ as $x_{n}=na$ and$\
g(x_{n}-x_{j})$ as $M_{nj}=e^{-(n-j)^{2}a^{2}/w^{2}}=\gamma
^{(n-j)^{2}}$, where $n,j$ are integers between $1$ and $N$. For large
enough positive integers $m$, $\gamma ^{m}\ll 1$ , so we can always
cut off at certain $m$ and set terms $\mathcal{O}\left( \gamma
  ^{m+1}\right) $ in $M_{nj}$ to zero.  The resulting $M_{nj}$ is then
a Toeplitz band matrix with bandwidth $2m+1$ \cite{trench}. The
solution $f_{ji}$ contains several exponential decay components with
different decay constants (see appendix for details), but $\left\vert
  -\gamma \right\vert $ characterizes the largest decay constant and
in the limit of large $\left\vert j-i\right\vert $ a single term wins
out with $f_{ji}\equiv f(x_{j}-x_{i})\sim (-\gamma )^{\left\vert
    j-i\right\vert }$. Numerical solution of the matrix equation
confirms this (see Fig~\ref{fig:f0_fx_two_panel}(a)).  An important
implication of this result is that we can set a truncation tolerance
error $%
\epsilon $ and only apply correction beams to those qubits with
$\left\vert f_{ji}\right\vert >\epsilon $. That will require about
$2\log\epsilon /\log\gamma =2(w/a)^{2}\log(1/\epsilon )$ beams,
independent of the system size. We expect this qualitative behavior to
persist for any beam profile that decays quickly with the increase of
distance from its center.

The amplitude $f(0)$, characterizing the required laser power, is
plotted in Fig.~\ref{fig:f0_fx_two_panel}(b) as a function of $w/a$
from exact numerical solution of Eq. (1). When $w/a\lesssim 1,$ $%
\gamma $ is small and from a truncation of Eq. (2) $g(k)\approx
1+2\gamma \cos(k)$, we find $f(0)\approx \left( 1/2\pi \right)
\int_{-\pi }^{\pi }dk\left[ 1/g(k)\right] \approx 1/\sqrt{1-4\gamma
  ^{2}}$. In the other region with $w/a\gtrsim 1$, the summation in
Eq. (3) can be approximated with an integration, which yields
$g(k)\approx \sqrt{\pi w^{2}/a^{2}}e^{-k^{2}w^{2}/\left( 4a^{2}\right)
}$ and therefore $%
f(0)\approx \frac{2a^{3}}{\pi ^{5/2}w^{3}}e^{\pi ^{2}w^{2}/\left(
    4a^{2}\right) }$. These two analytic expressions, also drawn in
Fig.~\ref%
{fig:f0_fx_two_panel}(b) agree well with the exact solution in their
respective regions. Note that for $w/a\lesssim 1$, $f(0)$ is close to
unity and the cost in the laser power in negligible. For $w/a\gtrsim
1$, $f(0)$ increases exponentially with $w^{2}/a^{2}$, and the scheme
becomes impractical when $%
w^{2}/a^{2}\gg 1$.  Our scheme is most effective in the region
$w/a\sim 1$, where it allows a reduction of the crosstalk error by
several orders of magnitude with just a few correction beams while
keeping the cost in the laser power still negligible.

The above analysis extends straightforwardly to higher dimensional
systems. Moreover, neither the assumption of homogeneous spacing nor
that correction beams center around each qubit is essential. We can
always treat the qubits as equidistant if we effectively modify the
beam profile $g(x-x_i)$ or $M_{nj}$ according to the actual qubit
spacings and the focus positions of the correction
lasers. 
For multi-qubit operations, the relative overhead of spatial
refocusing usually becomes lower. For instance, the quantum simulation
of arbitrary Ising interaction with $N$ trapped ion qubits requires
$N^{2}$ well focused laser beams in Ref. \cite{QS}. Without perfect
focusing, using the scheme here we still only need $%
N^{2}$ beams.

\section{Spectral refocusing}
Instead of using localized beams, an alternative for spatial
refocusing is to spectrally decompose the desired amplitude profile
and use broad beams of travelling plane waves with varying
wave-vectors $k$ to reconstruct a focused beam. Note here we do not
use light beams with different frequencies. We simply tilt the
traveling wave direction so that the effective spatial periodicity is
varied along the system axis. The desired spatial profile
$G(x_{j}-x_{i})=\delta _{ij}$, transformed to the momentum space, is a
constant function. For $N$ qubits, one can use $N$ plane waves with
$k$ evenly spread in the Brillouin zone $%
\left[ -\pi /a,\pi /a\right] $ to reconstruct the profile $\delta
_{ij}$. We may tilt a travelling wave with a fixed $k$ by different
angles with respect to the qubit array to get a varying wave-vector
component $k_{x}$ along the axis. For ion qubits in a harmonic trap,
the spacing is inhomogeneous and the exact amplitudes of the
components are not even, but can be obtained using the matrix
formalism of Eq. (1). For the plane wave with wave vector $%
k_{x}^{j}$, the amplitude at position $x_{n}$ is $M_{nj}=\exp
(i\,k_{x}^{j}\,x_{n})$. To get a perfectly focused beam at position
$x_{i}$, the amplitude $f_{ji}$ for the $k_{x}^{j}$ component is given
by the solution of the matrix equation $\sum_{j}M_{nj}\,f_{ji}=\delta
_{ni}$. The maximum $k_{x}^{j}=k\sin (\theta _{m})$ needs to be
comparable with$\,\pi /a$%
, so we require the laser angle is tunable over a window $[-\theta
_{m},\,\theta _{m}]$, where $\theta _{m}\approx \sin(\theta
_{m})\approx \pi /ka$ is typically small. For instance, in an ion trap
quantum computer, the ion spacing is about $5\mu m$ and the laser has
wavelength about $0.4\mu m$, which gives $\theta _{m}\sim 0.04\sim
2.3^{\circ }$. In Fig.~\ref%
{fig:The-amplitude-profile}(a), we show the amplitude distribution
$f\left( k_{x}^{j}\right) $ for $21$ ions in a harmonic trap and the
associated profile $G(x)$, which is basically a $\delta $-function at
ions' positions albeit with small wiggles at other location. This
spectral decomposition approach is particularly convenient for quantum
simulation where we need to simultaneously apply focused laser beams
on each ion \cite{QS}. With spectral decomposition, we only need to
apply a number of broad plane wave beams that cover all the ions, with
their angles tunable in a small window $%
[-\theta _{m},\,\theta _{m}]$.

\begin{figure}[tbp] \centering
\includegraphics[width=0.5\textwidth]{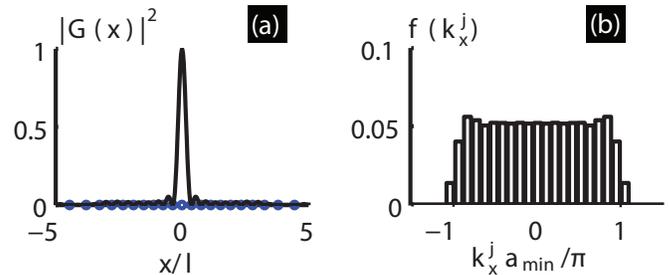}
\caption{(Color online) (a) Intensity (amplitude modulus squared)
profile of superposition of $21$ plane waves with different wave
vector components $k_{x}^{j}$ along the chain. Blue circles indicate
ions' positions. The unit of position $x$ is $l=\left( \frac{Z^2
e^2}{4\protect\pi \protect\epsilon_0 M
\protect\omega_z^2}\right)^{1/3}$, where $Ze$ and $M$ are the charge
and mass of each ion, $\protect\epsilon_0$ the free space permittivity
and $ \protect\omega_z$ the trap frequency along z axis. (b)
Amplitudes of spectral components.  Here $a_{min}$ is the smallest
spacing of ions in the middle of the chain.  }
\label{fig:The-amplitude-profile}
\end{figure}

\section{Application example}
As an example of application, we consider two-qubit quantum gates in
an ion chain. With spatial refocusing, we can perform high fidelity
entangling gates even when the Gaussian beam width is comparable with
the ion spacing, which significantly simplifies the experimental
realization. For two qubit operations, we need to illuminate only two
target ions in the chain. To be concrete, we consider a conditional
phase flip (CPF) gate $%
U_{jn}^{\text{CPF}}=\exp(i\pi \sigma _{j}^{z}\sigma _{n}^{z}/4)$
mediated by transverse phonon modes based on the scheme in
Refs. \cite{Zhu_prl,zhu_epl}. Here we only list the essential formula
and for detailed derivation we point the readers to the original
papers. From a practical point of view, one only needs to have
Eq.~\ref{eq:U} below in hand to understand this example. We define the
trap axis to be the $z$-direction. The gate is achieved by applying a
state-dependent ac-Stark shift on the ions, induced by a pair of Raman
beams with frequency detuning $\mu $ and wave vector difference $%
\Delta \mathbf{k}$ along the transverse direction $x$. The effective
Hamiltonian for the laser-ion interaction is $H=\sum_{j=1}^{N}\hbar
\Omega _{j}\cos\left( \Delta k\cdot q_{j}+\mu t\right) \sigma
_{j}^{z}$ where $q_{j}$ is the $j$-th ion's displacement operator
along $x$-direction and $\sigma _{j}^{z}$ acts on the qubit space of
the $j$-th ion. Expanding $q_{j}$ with normal phonon modes
\cite{DFV_James} $q_{j}=\sum_{k}b_{j}^{k}\sqrt{\hbar /2M\omega
  _{k}}(a_{k}+a_{k}^{\dagger })$ and assuming Lamb Dicke regime $%
\eta _{k}=\left\vert \Delta k\right\vert \sqrt{\hbar /2M\omega
  _{k}}\ll 1$, the interaction picture Hamiltonian under the rotating
wave approximation is $H_{I}=-\sum_{j,k=1}^{N}\hbar \chi
_{j}(t)g_{j}^{k}\left( a_{k}^{\dagger }e^{i\omega
    _{k}t}+a_{k}e^{-i\omega _{k}t}\right) \sigma _{j}^{z}$, where $%
g_{j}^{k}=\eta _{k}b_{j}^{k}$, $\chi _{j}(t)=\Omega _{j}(t)\sin\left( \mu
  t\right) $, $b_{j}^{k}$ is the normal mode wavefunction, $M$ is the
ion mass, and $%
\omega _{k}$ is the frequency of the $k$th motional mode. The
associated evolution operator is \cite{Zhu_prl,zhu_epl}
\begin{equation}
U(\tau )=\exp\left( i\sum_{j}\phi _{j}(\tau )\sigma _{j}^{z}+i\sum_{j<n}\phi
_{jn}(\tau )\sigma _{j}^{z}\sigma _{n}^{z}\right) , \label{eq:U}
\end{equation}%
where 
\begin{eqnarray*}
\phi_{j}(\tau) & = & \sum_{k}\left(\alpha_{j}^{k}(\tau)a_{k}^{\dagger}+h.c.\right)\\
\alpha_{j}^{k}(\tau) & = & \int_{0}^{\tau}\chi_{j}(t)g_{j}^{k}e^{i\omega_{k}t}dt\\
\phi_{jn}(\tau) & = & 2\int_{0}^{\tau}dt_{2}\int_{0}^{t_{2}}dt_{1}\times\\
& &\sum_{k}\chi_{j}(t_{2})g_{j}^{k}g_{n}^{k}\chi_{n}(t_{1})\sin\left[\omega_{k}(t_{2}-t_{1})\right]
\end{eqnarray*}
This is the key equation of this gate example so let us give more
comments to clarify the picture. The evolution operator contains
single-spin and two-spin part. The coefficients of the single-spin
part $\phi_j (\tau)$ are operators acting on the motional degree of
freedom. They give ions an internal state dependent displacement of
the motion. This would entangle the spin and motional degrees of
freedom. Since we care only about the spin part without measuring the
motional states, spin-motion entanglement reduces the purity of the
spin states. To get a high fidelity gate we desire a vanishing
single-spin part. The coefficients of two-spin part of evolution
$\phi_{jn}(\tau)$ are c-numbers and they only add a phase to the
state. Both $\phi_j (\tau)$ and $\phi_{jn}(\tau)$ can be tuned by
varying the Raman detuning $\mu$, the gate time $\tau$, and the
time-dependent laser Rabi frequency $\Omega_j(t)$. By definition of a
controled phase flip gate between ion $j$ and $n$, one should have
$\phi _{jn}(\tau )=\pi /4 $ with every other single-spin and two-spin
coefficient being zero. To perform such a gate, we shine lasers to
ions $j$ and $n$ only, i.e. $\Omega_i=0$ for $ i\ne j, n$, and
optimize over $\mu$ so that the effective evolution best approximates
$U_{jn}^{\text{CPF}}$. For simplicity, here we assume a time
independent $\Omega$ and pick a relatively long gate time $\tau
=180\tau _{0}$ ($\tau_{0}\equiv 2\pi /\omega _{z}$ is the trap
period). The gate fidelity is quantified by $F=Tr_{m}\left\langle \Psi
  _{f}\right\vert U(\tau )\left\vert \Psi _{0}\right\rangle
\left\langle \Psi _{0}\right\vert U^{\dagger }(\tau )\left\vert \Psi
  _{f}\right\rangle $, where $%
\left\vert \Psi _{0}\right\rangle =\frac{1}{2}\left( \left\vert
    0\right\rangle +\left\vert 1\right\rangle \right) \otimes \left(
  \left\vert 0\right\rangle +\left\vert 1\right\rangle \right) $ is
the assumed initial state, $\left\vert \Psi _{f}\right\rangle \equiv
U_{jn}^{\text{CPF}}\left\vert \Psi _{0}\right\rangle $ is the ideal
final state and $Tr_{m}$ indicates tracing over all the motional
modes.

\begin{figure*}[tbp]
\centering
\includegraphics[width=0.98\textwidth]{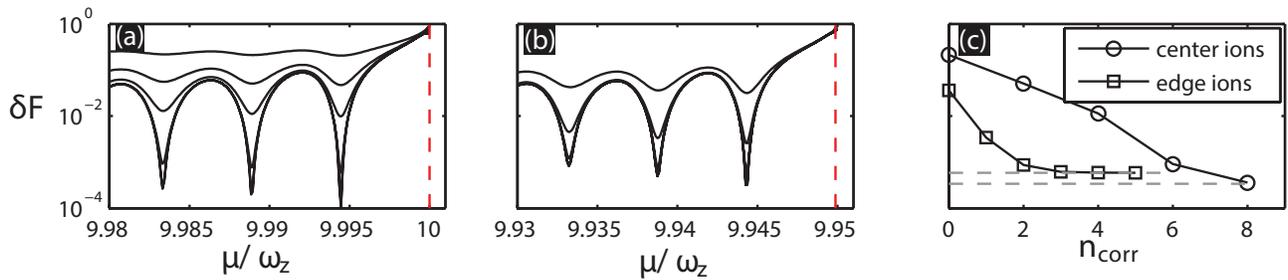}
\caption{(Color online) Fig (a), (b): Infidelity ($\protect\delta F \equiv
1-F$) of the CPF gate vesus the Raman detuning $\protect\mu$ for (a) two
ions in the center and (b) two ions on one edge in a $20$-ion chain.
Vertical dashed lines indicate the position of the transverse phonon modes.
The beam waist is set to $15\%$ larger than the minimum spacing (at the
center) of ions and about $2/3$ of the maximum spacing (at the edge) of the
chain. In (a), curves from top to bottom are for the cases with $0, 2, 4, 6, 8$
correction beams, respectively; in (b), curves from top to bottom are for
the cases with $0, 1, 2, 3$ correction beams. Fig. (c): Infidelity under a
fixed Raman detuning $\protect\mu=9.9888\protect\omega_{z}$ for center ions
and $\protect\mu=9.9387\protect\omega_{z}$ for edge ions, as a function of
the number of correction beams $n_{\text{corr}}$. Dashed lines denote the infidelity under
perfect focusing (with zero crosstalk error). }
\label{fig:Fid_detuning}
\end{figure*}

Similar to real experiments, we apply Gaussian beams to the target
ions $j,n$. We consider two entangling CPF gates in a $20$-ion chain
with $%
\omega _{x}/\omega _{z}=10,$ one for two center ions and the other for
two ions on one edge, with the beam width about $15\%$ larger than the
separation of the two center ions and $2/3$ of separation of the two
edge ions. The ion spacings and laser beam width are fixed throughout
the calculation. Clearly the condition $w/a\ll 1$ is violated in both
cases.  All the transverse phonon modes are assumed to be initially in
thermal states with the same temperature $T$ such that the center of
mass mode has one phonon on average, a typical situation after Doppler
cooling. We scan over the Raman detuning $\mu$ and for each $\mu$
optimize over $\Omega_j$ and $\Omega_n$ to find the best possible gate
fidelity. As expected, without applying correction beams the fidelity
of the gate is rather low (see the top curves in
Fig.~\ref{fig:Fid_detuning} (a) and (b)). However, keeping all other
parameters fixed, the gate error is largely reduced by including only
one correction beam and including two correction beams the fidelity
gets very close to the ideal case. For the center ions, three
correction beams on both sides already reduce the gate error by nearly
three orders of magnitude. As shown in Fig~\ref{fig:Fid_detuning}(c),
the gate infidelity (t1-fidelity) caused by the crosstalk error decreases
exponentially with the number of correction beams, until one
approaches the optimal value set by other error sources. Note that
with time constant $\Omega_j$ and $\Omega_n$, there is an intrinsic
gate fidelity due to the lack of control knobs, shown in
Fig~\ref{fig:Fid_detuning}(c) as dashed
lines.  

\section{Experimental implementation and error resistance}

The proposed spatial refocusing technique is ready to implement in
many quantum computation architectures, such as harmonically trapped
ion crystals \cite{qsim, Lin_epl} or arrays of
micro-traps \cite{array_microtraps}, Rydberg atoms in optical lattices
\cite{rydberg_lattice}, arrays of optical tweezers
\cite{tweezer_array}, etc.  After measurement of qubit positions,
laser focusing positions, and the laser beam profile, one only needs
to apply the inverse linear transformation $%
M_{nj}^{-1}$ to the target beam profile $G_{j}$ and use the result as
input to the beam delivery device. Removing the need of strong
focusing, this scheme should significantly simplify the required
optics. Another nice feature is that we do not even require each beam
to center at each qubit, as long as the beam positions are known and
fixed. The scheme requires coherence between the correction
beams. Since Raman beams are used we only need to stabilize the
relative phase between the Raman beams. We also note that in small
scale systems, the $\log(1/\epsilon)$ scaling of the number of
required correction beams $n_{\text{corr}}$ with truncation error is often
irrelevant. An array of $N$ coherent pulses should always suffice for
the generation of arbitrary laser strength profile for $N$ qubits. So
one would never need 10 beams to address 5 qubits.

\begin{figure}[tbp] \centering
\includegraphics[width=0.45\textwidth]{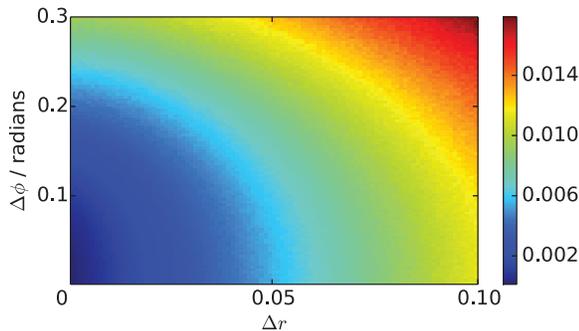}
\caption{(Color online) Average intensity error $\bar{\epsilon}$ as a
  function of standard amplitude/phase error $\Delta r$/$\Delta \phi$. The
  color encodes value of $\bar{\epsilon}$. Each point is obtained with
  5000 random samples of $r_j\sim\mathcal{N}(0, \sigma^2=\Delta r^2)$ and
  $\phi_j\sim\mathcal{N}(0, \sigma^2=\Delta \phi^2)$. }
\label{fig:error-plot}
\end{figure}

In practice, spatial refocusing is subject to several types of
experimental noise. First of all, the ions are not stationary point
particles. Their positions fluctuate thermally and quantum
mechanically. Second, the amplitudes and phases of each beam in the array may deviate from the prescription.  It is unclear whether the interference is robust to these deviations. We first estimate the position fluctuations of the ions. Take a 21 ion chain as example, the ion spacing
vary between $1.02 \upmu \text{m}$ and $1.78 \upmu \text{m}$ with the
smallest spacing in the middle of the chain. Among the axial motional
modes the center of mass mode has the lowest frequency, about $2\pi
\times 1 \text{MHz}$ and the corresponding oscillator length is
$\sqrt{\hbar/2M\omega_z}\approx 5.4\text{nm}$. The other axial modes
all have higher frequencies and the oscillator lengths are even
smaller. Assuming the Doppler cooling limit, i.e. with temperature
given by $k_B T=\hbar \Gamma/2$ and the cooling transition linewidth
$\Gamma \approx 2\pi\times 20 \text{MHz}$, the center of mass mode
along $z$ contains on average $\frac{k_BT}{\hbar \omega_z}\approx 10$
phonons for a trap with $\omega_z=2\pi \times 1\text{MHz}$. With these
realistic data, exact numerical calculation taking all the axial modes
into account shows that for each ion the standard deviation of
position ranges from $6.5 \text{nm}$ to $10\text{nm}$, at least two
orders of magnitude smaller than the inter-ion spacing. So for our
purpose here the noise caused by ions' thermal motion is
negligible. For the second problem, since the laser beams superpose
linearly to give the final refocused pulse, an arbitrary deviation of
the j-th pulse's amplitude $\delta f(x_j -x_i)$ only add noise $\delta
f(x_j -x_i)g(x-x_j)$ to
the final amplitude distribution $G(x-x_i)$. To consider both
strength and phase error of the laser, we allow the deviation $\delta
f(x_j -x_i)$ to be a complex number. To quantify the effect of $\delta
f(x_j -x_i)$, we parametrize the deviation as follows
\begin{equation}
f(x_j -x_i) + \delta f(x_j -x_i) = f(x_j -x_i) (1 + r_j) \exp(i \phi_j)
\end{equation} 
where the real numbers $r_j$ and $\phi_j$ measure respectively the
relative amplitude error and phase error of the beam on ion $j$.  Each
$r_j$/$\phi_j$ is sampled from the normal distribution with zero mean
and standard deviation $\Delta r$/$\Delta \phi$, i.e. $r_j\sim\mathcal{N}(0, \sigma^2=\Delta r^2)$ and  $\phi_j\sim\mathcal{N}(0, \sigma^2=\Delta \phi^2)$.  We define the
quantity
\begin{equation}
\epsilon =
\frac{1}{N}\sum_j\left|\left|G(x_j-x_i)\right |^2 -
  \left|\tilde{G}(x_j-x_i)\right |^2\right|
\end{equation}
to measure the difference of actual and ideal intensity
distribution. We now do a numerical simulation to investigte the
robustness of the interference. We take a 21-ion chain harmonically
trapped and try to address the central ion, $i=11$. The ideal target
is $G(x_j-x_{11})=\delta_{j,11}$. Assume the addressing  beams have a Gaussian profile with width the same as
the distance between 11-th and 12-th ion. We randomly sample $r_j$ and
$\phi_j$ 5000 times, calculate $\epsilon$ for each sample and plot the
average $\bar{\epsilon}$ as a function of $\Delta r$ and $\Delta
\phi$, in Fig~\ref{fig:error-plot}. We found that the interference
pattern is pretty robust. For 5\% standard amplitude error and 0.2
radians phase error, the average intensity error $\bar{\epsilon}$ is
still below 1\%. In terms of gate infidelity, we did numerical experiments and found that 1\% intensity error induces on the order of $10^{-2}$ ($10^{-3}$) infidelity for two center ions with $n_{\text{corr}} = 8 $ (edge ions with $n_{\text{corr}}  = 5$), with every other parameter the same as described in caption of fig~\ref{fig:Fid_detuning}. For 0.5\% intensity error, both infidelities are on the $10^{-3}$ level.

\section{Conclusion}
In summary, we have proposed a spatial refocusing technique to achieve
effective individual addressing and reduce crosstalk error in a
general multi-qubit platform. The scheme is efficient as the crosstalk
error decreases exponentially with the number of correction beams, and
the cost in the laser power is modest even when the beam width is
comparable with the qubit separation. The scheme works universally for
any type of quantum gates and can apply to any quantum computational
platform.

\section*{Acknowledgements}
This work was supported by the NBRPC (973 Program) 2011CBA00300
(2011CBA00302), the IARPA MUSIQC program, the ARO and the AFOSR MURI
program, and the DARPA OLE program. We thank Kihwan Kim for helpful
discussion.

\section*{Appendix. Solving the envelope function with Toeplitz matrix theory}
According to the Toeplitz matrix theory, the general solution to the
equation $M_{lj}f_{ji}=\delta _{li}$ has the form $f_{ji}=%
\sum_{k}c_{k}^{+/-}\left( a_{k}\right) {}^{j-i}$ where $c_{k}^{+}$ and
$%
c_{k}^{-}$ are for the regions $j>i$ and $j<i$, respectively. Here
$a_{k}$ are the roots of the polynomial $P_{n}(x)=x^{n}\left(
  1+\sum_{m=1}^{n}\left( 1\left/ x^{m}\right. +x^{m}\right) \gamma
  ^{m^{2}}\right) $ and $c_{k}^{+/-}$ are coefficients to be
determined (the band-width of the matrix $M_{lj}$ is $%
2n+1$). Our first observation is that the roots come in pairs ($a$ ,
$1/a$ ) due to the symmetry $x\leftrightarrow 1/x$. Thus $f_{ji}$ is
composed of terms like $c_{k}^{+/-}(a_{k})^{j-i}$ decaying
(increasing) exponentially with $\left\vert j-i\right\vert $ if
$a_{k}<1$ ($a_{k}>1$). In the region $%
j>i$ ($j<i$) , boundary condition at $\left\vert j-i\right\vert
\rightarrow \infty $ requires $c_{k}=0$ for $a_{k}>1$
($a_{k}>1$). Note that in the large $\left\vert j-i\right\vert $
limit, the $a_{k}$ closest to the unity should dominate since other
components die out more quickly. Next we prove that $-\gamma $ (and
hence $-1/\gamma $) is a root of $P_{n}(x)$ when $n$ is sufficiently
large.

\begin{eqnarray*}
\frac{P_{n}(-\gamma )}{(-\gamma )^{n}} &=&1+\sum_{m=1}^{n}(-1)^{m}\left(
\gamma ^{m^{2}+m}+\gamma ^{m^{2}-m}\right) \\
&=&1+\sum_{m=1}^{n}(-1)^{m}\gamma
^{m^{2}+m}+\sum_{m=0}^{n-1}(-1)^{m+1}\gamma ^{m^{2}+m} \\
&=&(-1)^{n}\gamma ^{n^{2}+n}+\sum_{m=1}^{n-1}(-1)^{m}(1-1)\gamma ^{m^{2}+m}
\\
&=&(-1)^{n}\gamma ^{n^{2}+n}\rightarrow 0,\;\text{when }n\text{ is large.}
\end{eqnarray*}

The characteristic quantities of $P_{n}(x)$ are $\gamma $, $\gamma
^{4}$, $%
\gamma ^{9}$, ..., of which the one closest to the unity is $\gamma
$. This leads us to conjecture $(-\gamma )$ is the root of $P_{n}(x)$
closest to $1$ in magnitude. This turns out to be true. Since
$\tilde{P}%
_{n}(x)=P_{n}(x)/x^{n}>0$ when $x>0$, there is no positive root. Let
us focus on the interval $[-1,0)$. For $n=1$,
$\tilde{P}_{1}(x)=1+\gamma (1/x+x) $ is monotonically decreasing from
$\tilde{P}_{1}(-1)=1-2\gamma $ to $\tilde{P}_{1}(0^{-})\rightarrow
-\infty $ and there is one root in this interval:
$\frac{-1+\sqrt{1-4\gamma ^{2}}}{2\gamma }\approx \frac{%
  -1+1-2\gamma ^{2}}{2\gamma }=-\gamma $. When increasing $n$ by $1$,
we include one more term $Q_{n+1}(x)=\left( 1/x^{n+1}+x^{n+1}\right)
\gamma ^{(n+1)^{2}}$. Due to the small factor $\gamma ^{(n+1)^{2}}$,
the contribution of $Q_{n+1}$ can be comparable with that of $Q_{n}$
only when $%
\left\vert x\right\vert \lesssim \gamma ^{2n+1}$. Since $Q_{n}(0^{-})$
approaches $+\infty $ for even $n$ and $-\infty $ for odd $n$ and
$Q_{n}$ is always monotonic on $[-1,0)$, adding one more term always
introduces one more turning point in $\tilde{P}_{n}(x)$ and thus adds
one more root with magnitude much smaller than the previous
roots. Therefore $(-\gamma )$ is the root with the largest magnitude
by far on $[-1,0)$. We therefore conclude $f_{ji}\propto (-\gamma
)^{|j-i|}$ when $\left\vert j-i\right\vert $ is large.


\end{document}